\documentclass[a4paper,english]{article}
\usepackage[T1]{fontenc}
\usepackage[latin1]{inputenc}

\makeatletter

\providecommand{\LyX}{L\kern-.1667em\lower.25em\hbox{Y}\kern-.125emX\@}
 \newcommand{\lyxaddress}[1]{
   \par {\raggedright #1 
   \vspace{1.4em}
   \noindent\par}
 }
 \newenvironment{lyxlist}[1]
   {\begin{list}{}
     {\settowidth{\labelwidth}{#1}
      \setlength{\leftmargin}{\labelwidth}
      \addtolength{\leftmargin}{\labelsep}
      }}
   {\end{list}}

\usepackage{babel}
\makeatother
\begin{document}

\title{Suppression of superconductivity due to spin imbalance in Co/Al/Co
single electron transistor}

\author{Jan Johansson, Mattias Urech, David Haviland, and V. Korenivski}

\maketitle

\lyxaddress{Nanostructure Physics, KTH, AlbaNova University Center, 10691 Stockholm}

\begin{abstract}
Transport properties of ferromagnetic/non-magnetic/ferromagnetic single
electron transistors are investigated as a function of external magnetic
field, temperature, bias and gate voltage. By designing the
magnetic electrodes to have different switching fields, a two-mode
device is realized having two stable magnetization states, with the
electrodes aligned in parallel and antiparallel. Magnetoresistance of approximately 100\%
is measured in Co/AlO$_{X}$/Al/AlO$_{X}$/Co double tunnel junction
spin valves at low bias, with the Al spacer in the superconducting
state. The effect is substantially reduced at high bias and temperatures
above the $T_{C}$ of the Al. The experimental results are interpreted
as due to spin imbalance of charge carriers resulting in suppression
of the superconducting gap of the Al island.
\end{abstract}

\section{Introduction}

Electron tunnelling from a ferromagnet into a normal metal or a superconductor
results in a non-equilibrium spin population persisting over a characteristic
distance known as the spin diffusion length, typically 10-100 nm at low temperatures \cite{m-t,mark}.
With the recent advances in nano-fabrication techniques it has become
possible to study structures with dimensions on the same length scale,
where spin coherence and relaxation effects play an important role.
One of the implementations of a spin transport device is a double
junction with ferromagnetic outer electrodes that can be magnetically
switched to align in parallel (P) or antiparallel (AP). The spin imbalance
and accumulation on the superconducting island in the AP configuration
is expected to produce large conductance variations by suppressing
the gap \cite{takahashi}. Recently reported experiments on Co/Al/Co
double tunnel junctions \cite{chen-x,chen-prl} have indeed been interpreted
in terms of suppressed superconductivity of Al in the magnetic AP
state of the device. Interestingly, the magnitude of the magnetoresistance
(gap suppression) was found to strongly depend on the sweep rate of
the external magnetic field \cite{chen-x}, suggesting that the AP
state of the device was unstable ({}``thermal activated'') on the
experimental time scale of seconds to minutes. We show in this paper
that this reported \emph{sweep-rate-dependent} magnetoresistance (MR)
in Co/Al/Co double tunnel junctions is not connected with the relative magnetic alignment of the two ferromagnetic electrodes. By carefully designing the magnetic electrodes of the structure
we are able to achieve a well controlled and stable AP to P switching,
which results in a pronounced and \emph{field-sweep-rate-independent}
MR. Similar to \cite{chen-x,chen-prl} we observe a sweep-rate-dependent
contribution to the MR of the devices, which we find to be unrelated
to the magnetization reversal in the Co electrodes.

\section{Experimental details}

The structures, consisting of an Aluminium island separating two Cobalt
electrodes as shown in Fig. 1, were fabricated using e-beam lithography
and the two-angle shadow evaporation technique \cite{dolan}. A 15
nm thick Al layer was deposited on oxidized Si and subsequently in-situ
oxidized in 100 mTorr of O$_{2}$ prior to deposition of 40 nm thick,
60 and 70 nm wide Co electrodes spaced by $\sim 400$ nm. The difference
in width resulted in different magnetostatic shape anisotropy, which
in turn determined the switching field of the electrodes. The orientation
and length of the Co fingers (extending past the Al island) was chosen
so as to minimize the stray fields in the Al due to the open ends,
promote the AP magnetostatic coupling between the fingers as well
as minimize magnetization curling in the junction area \cite{urech}. These considerations
were important in achieving a stable AP magnetic state of the device.
The external quasi-static field was applied along the Co electrodes,
perpendicular to the longer side of the Al island. MR and I-V characteristics
were measured at temperatures ranging from 250 mK to above the $T_{C}$
of Al ($\sim 1.2$ K).

\section{Results and discussion}

Fig. 2 shows the resistance of a typical double-junction as a function of external field swept at two different rates, 4 and 15 Oe/s. The decrease in the resistance at high
fields is caused by the direct influence of the external field, which
suppresses the superconducting gap of the Al and thereby enhances
the quasi-particle tunnelling. As the field is lowered (see the 15
Oe/s curve in Fig.2) the resistance increases and would be expected
to follow a bell-like shape before decreasing again at high negative
fields. However, a pronounced minimum is observed at a relatively
low field ($\sim $-150 Oe), which appears similar to Giant Magnetoresistance
in spin-valves or magnetic tunnel junctions \cite{modera}, and has previously
been interpreted as arising from magnetic P to AP switching \cite{chen-x,chen-prl}.
 This minimum is significantly reduced in magnitude (3-fold) as the sweep rate of the magnetic field is reduced to 4 Oe/s (upper curve in Fig.2) and eventually vanishes in quasi static field measurements (sweep rate < 1 Oe/s). Additionally, we observe a second, somewhat less pronounced minimum
at $\sim 1$ kOe. The second minimum, however, is essentially unchanged as the sweep
rate is reduced. Clearly, two effects must be at work here. We believe
that the interpretation of the sweep-rate dependent MR minimum (see
\cite{chen-x,chen-prl}) as arising from the P to AP switching of
the Co electrodes is faced with two large difficulties. Indeed, after
the sample has undergone a complete saturation the resistance versus
H levels off and starts to decrease sharply before the external field reverses direction,
where the field continues to favour a parallel alignment of the magnetic
electrodes (even more so in the geometry of ref. \cite{chen-prl}, where
the inter-electrode magnetostatic coupling favours the P state). Only when the sign of the external field is reversed one (or
both) of the electrodes can switch in order to minimize the Zeeman
energy. The switching field is then negative and determined by the
shape anisotropy of the electrode. Secondly, the measured time dependence
on the scale of seconds to minutes was argued to come from a slow magnetic
switching/relaxation of the Co electrodes \cite{chen-x}. This argument,
however, is off by at least 10 orders of magnitude since nano-magnets
are known to switch on the sub-nanosecond scale \cite{brown}. The
characteristic precessional time scale is set by the inverse of the
ferromagnetic resonance frequency, $f_{r}=\gamma \sqrt{4\pi M_{s}H_{a}}\sim 10$
GHz, where $4\pi M_{s}(Co)=16$ kG and the anisotropy field is $H_{a}\sim 1$
kG for our geometry. We obtain magnetic reversal times of the order of $10^{-10}$ seconds using
micromagnetic simulations for our electrode geometry. We therefore conclude, in contrast to \cite{chen-x,chen-prl},
that the \emph{sweep-rate-dependent} part of the observed MR is unrelated
to magnetic switching of the Co electrodes and must have a different
origin. We can only point out that the field scale at which the sweep rate dependent minimum of MR is observed is approximately 7 times lower in our case than of ref. \cite{chen-prl}. The location of the minimum scales
inversely with the cross-section of the Al island perpendicular to
the applied field. This implies that in these two experiments
the minimum in resistance is observed at roughly the same flux through
the Al spacer.

The \emph{sweep-rate-independent}  MR shows the field
dependence of a classical spin-valve (see Fig.2). After saturation
in a large positive field (P state) one of the Co electrodes (the
wider of the two, having weaker shape anisotropy) switches only when
the field is reversed to -800 Oe. The second (narrower, higher
shape anisotropy) electrode switches at -1000 Oe. In this 200 Oe field
window an AP state of the device is achieved, which results in a pronounced
and stable MR. Once the device is set in the AP state by ending the
sweep between 800 and 1000 Oe, it remains in it after removal of the
field, stabilised by the magnetic shape anisotropy of the Co strips.
The I-V characteristics measured in a so-prepared AP state together
with the I-V of P are shown in Fig.3. From the I-V characteristics in the P state and the response to a gate voltage, a good estimate of the sample parameters can be obtained: superconducting gap of $ \Delta_0 = 200\ \mu V$, junction capacitance of $ 0.57\ fF$, junction resistance $ 42.5\ k\Omega$ and gate capacitance of $ 1\ aF$. The gap in the AP state is clearly reduced. The spin accumulation on the Al island, due to
the spin-valve effect, results in Cooper pair braking which reduces
the gap thereby increasing quasi-particle tunnelling and thus in a lower resistance of the double tunnel junction \cite{takahashi}.

The MR, defined as the difference between the P and AP resistances normalized to the AP resistance, is shown in Fig.4 as a function of bias voltage. The three curves correspond to the temperatures 250 mK, 450 mK and 1200 mK. The resistance, R=V/I at fixed V, is obtained from the I-V data for the stable P and AP states, such as shown in Fig.3. The spin-valve effect is most pronounced at low bias of $\sim 200\ \mu V$, which approximately equals the superconducting gap. Above and below this voltage the effect is weaker, and asymptoticly goes to zero for large bias. As the temperature is increased the MR decreases. Above $T_{c}$ we see no MR indicating that the spin coherence length in the normal state is shorter than the distance between the two Co electrodes. The noise in the data at very low bias is due to numerical uncertainties in this range ($\frac{V\rightarrow 0}{I\rightarrow 0}$) and fluctuating background charges in the vicinity of the Al island. The measured behaviour is qualitatively consistent with the theoretical results of \cite{takahashi}. We observe a negative, bias dependent MR, which decreases with increasing temperature, see fig. 2 and 3 of ref. \cite{takahashi}. 

Thus, we have achieved a controlled P to AP switching in Co/Al/Co magnetic single electron transistors. In the AP state we observe a clear reduction of the superconducting gap due to the spin accumulation effect, which results in a maximum MR of close to 100\% at a voltage bias close to the superconducting gap. The MR vanishes at large bias and above the $T_{C}$ of the Al island.

\section*{Acknowledgement}

J. J. and M. U. gratefully acknowledge support from the Swedish SSF
under the graduate school for quantum devices.

\section*{Figure captions}

\begin{lyxlist}{00.00.0000}
\item [Fig.1]SEM image of a Co/Al/Co double-tunnel junction. 
\item [Fig.2]Magnetoresistance of a Co/Al/Co double-tunnel junction measured
with the field swept at 4 and 15 Oe/s. The 4 Oe/s curve is offset by +100 k$\Omega$. 
\item [Fig.3]I-V characteristics of a Co/Al/Co double-tunnel junction.
in the stable parallel (P) and antiparallel (AP) state of the Co electrodes.
\item [Fig.4]Magnetoresistance of a Co/Al/Co double-tunnel junction, $MR=(R_{AP}-R_{P})/R_{AP}$,
as a function of bias voltage.\end{lyxlist}

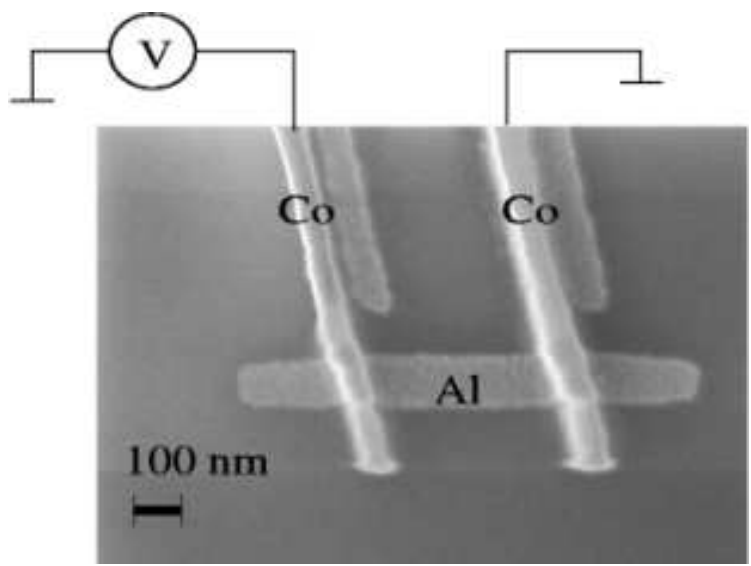
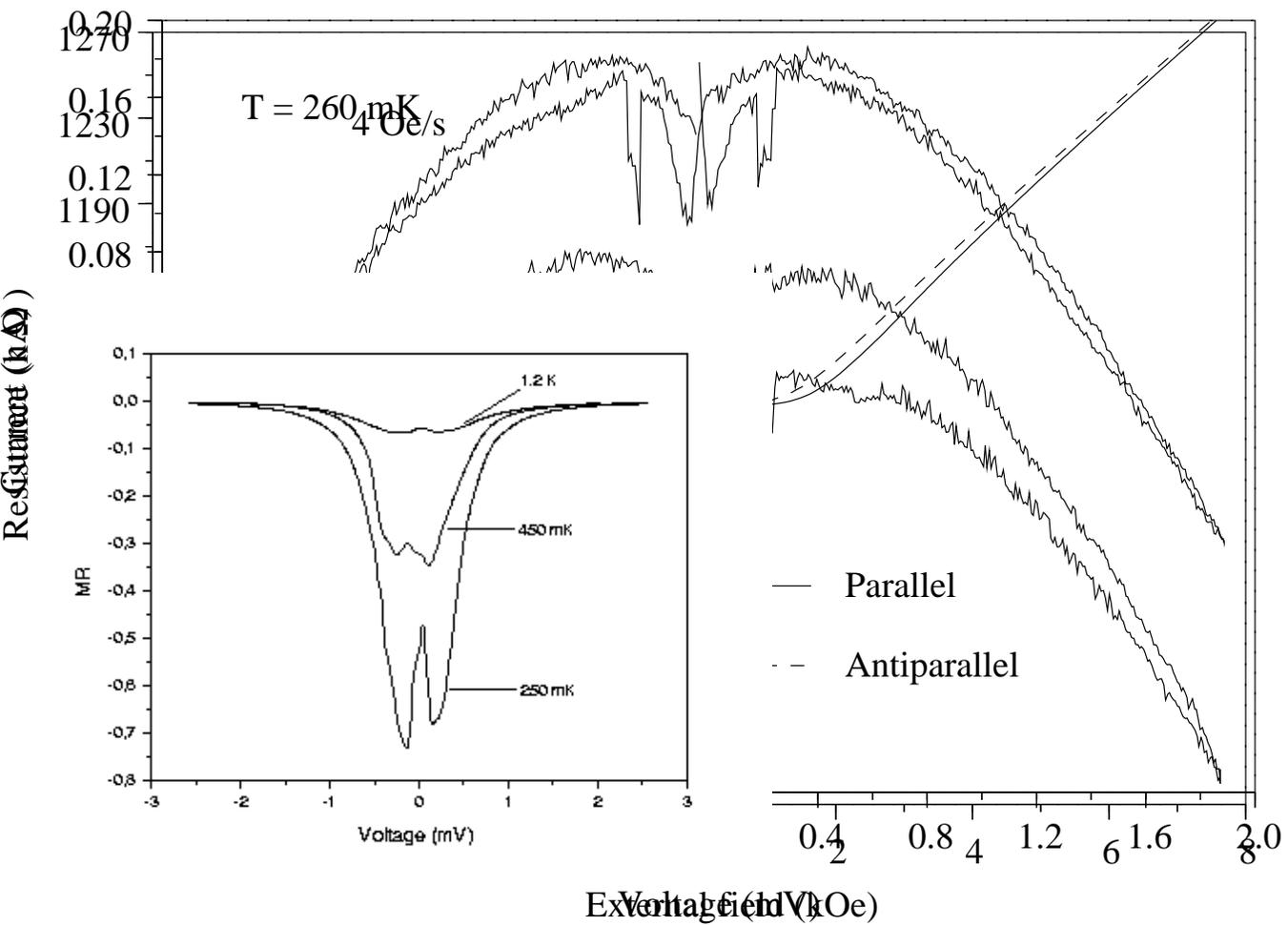
\end{document}